\documentclass[a4paper]{article}
\usepackage[T1]{fontenc}
\usepackage[utf8]{inputenc}
\usepackage{lmodern}
\usepackage[english]{babel}
\usepackage{amssymb,amsmath}
\usepackage{graphicx}
\usepackage{subfigure}
\usepackage{multirow,array}
\usepackage{epsfig}
\usepackage{authblk}
\usepackage{setspace}
\title{Fluid friction and wall viscosity of the 1D blood flow model}
\author[1]{Xiao-Fei WANG}
\author[2]{Shohei NISHI}
\author[2]{Mami MATSUKAWA}
\author[1]{Arthur Ghigo}
\author[3]{Pierre-Yves LAGR\' EE}
\author[1]{Jose-Maria FULLANA}
\affil[1]{Sorbonne Universit\'es, UPMC Univ Paris 6, UMR 7190, Institut Jean le Rond $\partial$'Alembert}
\affil[2]{Doshisha University, Department of Electrical Engineering,
Laboratory of Ultrasonic Electronics}
\affil[3]{CNRS, UMR 7190, Institut Jean le Rond $\partial$'Alembert}

\date{\today}

\begin{document}
\maketitle

\begin{abstract}

  We study the behavior of the pulse waves of water into a flexible
  tube for application to blood flow simulations.  In
  pulse waves both fluid friction and wall viscosity are damping
  factors, and difficult to evaluate separately.  In this paper, the
  coefficients of fluid friction and wall viscosity are estimated by
  fitting a nonlinear 1D flow model to experimental data.
  In the experimental setup, a distensible tube is connected to a
  piston pump at one end and closed at another end.  The pressure and
  wall displacements are measured simultaneously.  A good agreement
  between model predictions and experiments was achieved.
 For amplitude decrease, the effect of wall viscosity on the
  pulse wave has been shown as important as that of fluid viscosity.
 \footnote{Submitted to Journal of Biomechanics, 2015. (jose.fullana@upmc.fr)}
\end{abstract}




\section{Introduction}

Although the modeling of blood flow has a long history, it is still a
challenging problem.  Recently 1D modeling of blood flow circulation
has attracted more attention. One reason is that it is a well balanced
option between complexity and computational cost (see
e.g.~\cite{alastruey2012arterial, cousins2012boundary,
  olufsen2012rarefaction,reymond2013physiological,
  van2011pulse,yamamoto2011experimental}).  It is not only  very
important to predict the time-dependent distributions of flow
rate and pressure in a network,
but it is also important to be able to predict mechanical properties
of the wall (see~\cite{lombardi2014inverse}), it is clear that could help
the underestanding of cardiovascular pathologies.

The 1D fluid dynamical models are non nonlinear and are able to
predict flow, area and pressure. Within the dynamical system there
exist several damping factors, such as the fluid viscosity, the wall
viscoelasticity, the geometrical changes of vessels, etc.  Previous
studies have shown that in vessels without drastic geometrical
variations (i.e. no severe aneurysms or stenoses), the fluid viscosity
and wall viscoelasticity are the most significant damping
factors~\cite{matthys2007pulse}.  Comparisons between the 1D model and
\textit{in-vivo}
data~\cite{holenstein1980viscoelastic,reymond2009validation} suggest
that the predictions of a viscoelastic 1D model is significantly more
physiological than those of an elastic one which contains high
frequencies in the pulse which is not observed experimentally.
But the comparisons were
only qualitative or semi-quantitative due to the limited accuracy of
associated
non-invasive measurements and the lack of patient-specific parameter
values of the 1D model for each subject.

Quantitative comparisons can be done with \textit{in-vitro}
experimental setups.
Reuderink et al.~\cite{reuderink1989linear} connected a distensible
tube to a piston pump, which ejects fluid in pulse waves throughout the
tube, and the experimental data were compared against numerical predictions
of several formulations of the 1D model.  In the first formulation,
they proposed an elastic tube law and Poiseuille's theory to account for
the fluid viscosity, and their studies underestimated the
damping of the waves and predicted shocks,  not observed in
the experiments.  In another formulation, still linear, the fluid
viscosity was predicted from the Womersley theory with a
viscoelastic tube law which gave a better match between the
predictions and the experiments. A similar experiments setup was
proposed by  Bessems et
al.~\cite{bessems2008experimental}  using a 3-component Kelvin
viscoelastic model to model the wall behavior, however in this work, both
 the convective and
fluid viscosity terms were neglected.  Alastruey et
al.~\cite{alastruey2011pulse} presented a comparative study using an
experimental setup with a network, they measured the coefficients of a
Voigt viscoelastic model by tensile tests instead of fitting them
from the waves.  For the fluid viscosity term, they adopted a
value from literature, which was fitted from waves of coronary blood
flow with an elastic wall model~\cite{smith2002anatomically}.

In this paper, we study the friction and wall viscoelasticity using the
1D model and a similar experimental setup where pulse waves are
propagating in one distensible tube. However, there are three main
differences between our study and previous ones:

\begin{enumerate}

\item \emph{Both of the two damping factors (fluid friction and wall
    viscosity) are modelled.}  Although there are several theories to
  estimate the friction term (see,
  e.g.~\cite{bessems2007wave,lagree2000inverse,olufsen2000numerical}),
  the value is rarely determined experimentally besides the study of
  Smith et al. with an elastic model~\cite{smith2002anatomically}.  It
  is well known that fluid viscosity and wall viscoelasticity have
  damping influences on the pulse waves.  These slight differences are
  discussed in \cite{wang2014verification}, nevertheless it is
  difficult to evaluate them separately from pulse waves.  However,
  the viscoelasticity has smoothing effect on the waveforms whereas
  the fluid friction does not~\cite{alastruey2012physical}, we
  investigated this claim by
  only accounting for the amplitude or the sharpness of the signal.
The study shows the results of including both effects, one, or the
other.

\item \emph{The viscoelasticity of the wall is measured in a new
    manner.}  The viscoelasticity of a solid material is difficult to
  measure accurately, even in an \textit{in-vitro} setup.  In our
  study, the viscoelasticity is determined through the pressure-wall
  perturbation relation of the vessel under operating conditions.  The
  internal pressure is measured by a pressure sensor and the
  perturbation of the wall is measured by a Laser Doppler Velocimetry
  (LDV).

\item \emph{A shock-capturing scheme is applied as the numerical
    solver.}  In a nonlinear hyperbolic system, shocks may arise even
  if the initial condition is smooth (even for small viscoelasticity
  values).  The Monotonic Upstream Scheme for Conservation Laws
  (MUSCL) scheme is able to capture shocks without non-physical
  oscillations, and is applied to discretize the
  governing equations and compared to the MacCormack scheme.
\end{enumerate}

\section{Methodology}

\subsection{One-dimensional model}
We use the 1D governing equations for flows passing through an elastic
cylinder of radius $R$ expressed in the
dynamical variables of flow rate $Q$, cross-sectional area $A= 2 \pi R$ and
internal average pressure $P$. The 1D equations can be derived by the
integration over a cross-sectional area of the axy-symmetric Navier-Stokes equations
of an incompressible fluid at constant viscosity, giving the following
mass and momentum 1D conservation equations

\begin{eqnarray}
\frac{\partial A}{\partial t}+\frac{\partial Q}{\partial
  x}=0,  \label{eq:1DA} \\
\frac{\partial Q}{\partial t}+ \frac{\partial}{\partial x}(\alpha
\frac{Q^2}{A})+
\frac{A}{\rho}\frac{\partial P}{\partial x}=
-2\pi \nu \left [ \frac{\partial v_x}{\partial r} \right
  ]_{r=R}, \label{eq:1DQ}
\end{eqnarray}
where $v_x$ is the axial velocity, $\rho$ is the fluid density and
$\nu$ is the kinematic viscosity of the fluid. The parameter $\alpha$
and the last term, the viscous or drag friction, depend on the
velocity profile. In general, the axial velocity is also function of
the radius coordinate $r$, v.i.z. $v_x=v_x(r,x,t)$.  If we assume the
profile has the same shape $\Psi(r)$ in every vessel cross-section
along the axial direction, the velocity function can be separated as
$v_x=U(x,t)\Psi(r)$, being $U$ the average velocity.  If $\Psi(r)$ is
known, the parameter $\alpha$ and the derivative
$\frac{\partial v_x}{\partial r}$ that appears in the friction term
can be therefore calculated.  The friction drag can be
approximated by $-C_f Q/A$.  The radial profile $\Psi(r)$ is strongly
dependent on the Womersley number defined by $R\sqrt{\omega/\nu}$,
where the quantity $\omega$ is the angular frequency which
characterizes the flow.  If $\omega$ and $\nu$ are
approximately constant, only the radius $R$ influences $\alpha$ and
$C_f$, whose values should be determined by experiments for vessels
with various diameters.  When the transient inertial force is
large, the profile is essentially flat,
$\alpha=1$~\cite{smith2002anatomically}. With a thin
viscous boundary layer, the inviscid core and a no-slip
boundary condition, the friction term can be estimated (see
e.g.~\cite{bessems2007wave,olufsen2000numerical}).
When the transient inertial force is small, the profile is parabolic,
$\alpha=4/3$; the viscosity force is then dominating and $C_f=8\pi\nu$.
Using the power law profile proposed by Hughes and
Lubliner~\cite{hughes1973one}, Smith et
al.~\cite{smith2002anatomically} compute from coronary blood flow,
$C_f=22\pi\nu$ and $\alpha= 1.1$.  This value of $C_f$ is used on
other numerical
works~\cite{alastruey2011pulse,marchandise2009numerical} but setting
$\alpha=1$ for simplification.

The viscoelasticity of the wall can be described using different
viscoelastic models,
e.g.~\cite{holenstein1980viscoelastic,reymond2009validation,steele2011predicting}
with displaying disctint numerical
problems~\cite{raghu2011comparative,steele2011predicting}.  In this
study we use the two-component Voigt model, which relates the strain
$\epsilon$ and stress $\sigma$ in the equation
\begin{equation}
\sigma=E \epsilon+ \phi \frac{d \epsilon}{dt}, \label{Kelvin_Voigt_exp}
\end{equation}
where $E$ is the Young's modulus and $\phi$ is a coefficient for the
viscosity. In reference \cite{saito2011one,wanga2013effect} we have
shown that the model (i) fits
experimental data and (ii) it is able to filter high frequencies.

For a tube with a thin wall, the circumferential strain
$\epsilon_{\theta \theta}$ can be expressed as
\begin{equation}
\epsilon_{\theta \theta}=\frac{R-R_0}{(1-\eta^2)R_0}, \label{strain_exp}
\end{equation}
where $R_0$ is the reference radius without loading and $\eta$ is the
Poisson ratio, which is 0.5 for an incompressible material.  By
Laplace's law, the transmural difference between the internal pressure
$P$ and the external pressure $P_{ext}$ is balanced with the
circumferential stress $\sigma_{\theta\theta}$ in the relation
\begin{equation}
P-P_{ext}=\frac{h\sigma_{\theta\theta} }{\pi R}. \label{transmural_pressure_exp}
\end{equation}
Combining Eq.~\ref{Kelvin_Voigt_exp},~\ref{strain_exp} and \ref{transmural_pressure_exp}, we get
\begin{equation}
P-P_{exp}=\nu_e(R-R_0)+\nu_s\frac{dR}{dt}, \label{constitutive_exp}
\end{equation}
with
\[
\nu_e=\frac{Eh}{(1-\eta^2) A_0} \text{, and }
\nu_s=\frac{\phi h}{(1-\eta^2) A_0}.
\]
Note that the radius $R$ in the denominators of the two coefficients is
approximated by $R_0$ under the assumption that the perturbations are
small.

If we assume $P_{ext}$  constant and inserting
Eq.~\ref{constitutive_exp} into the 1D momentum equation to eliminate
$P$, gives
\begin{equation}
\frac{\partial Q}{\partial t}+ \frac{\partial}{\partial x}\bigl(\alpha \frac{Q^2}{A}+\frac{\beta}{3\rho}A^\frac{3}{2}\bigr)=
-C_f\frac{Q}{A} +C_v\frac {\partial ^2 Q}{\partial^2 x}, \label{momentumConserv_AQ}
\end{equation}
where
\[ \beta=\frac{\sqrt{\pi} Eh}{(1-\eta^2)A_0} \text{, and }
C_v=\frac{\sqrt{\pi}\phi h}{2\rho(1-\eta^2)\sqrt{A_0}}. \]

The 1D model was numerically solved by two approaches : MacCormack and
MUSCL. More details on the integration schemes and on the treatment of
the boundary condition are
in~\cite{delestre2013well,wang2014verification}. More precisely here
the boundary condition modeling the stainless rod in the experiment,
a total reflection boundary condition, can be numerically achieved by
imposing a mirror condition at the end of the elastic tube.

\subsection{Experimental setup}

\begin{figure}
\centering
\includegraphics[width=0.75\textwidth]{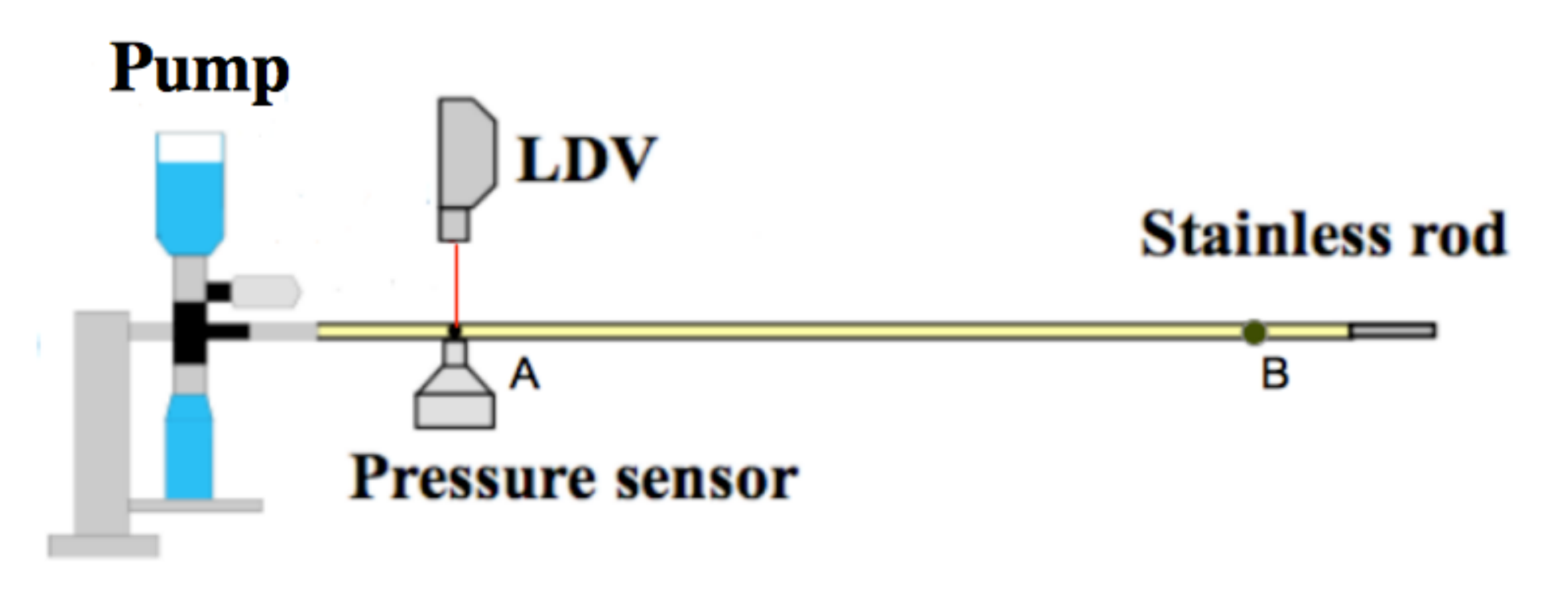}
\caption{Experimental setup : the elastic tube (in yellow) is closed
  by a stainless rod at the right end (in grey). The points $A$ and
  $B$ indicate the measurement sites. Parameters of the tube and fluid
  are summarized in Table~\ref{tab:Param_tube_fluid}.}
\label{fig:experimental_setup}
\end{figure}

The experimental setup is shown in Fig.~\ref{fig:experimental_setup}.
The piston pump (TOMITA Engineering) injects fluid (water) into a
polyurethane tube.  Theoutput of the pump is a sinusoidal function in
time, whose period and duration can be programmed through a computer.
At the measurement points, a pressure sensor (Keyence, AP-10S) is
inserted into the tube.  The perturbation of the tube wall is measured
by a LDV (Polytec, NLV-2500).  The pump, the pressure sensor and the
LDV are controlled by a computer, which synchronizes the operations of
the instruments and stores the measurement data at 10 $KHz$.  The end
of the tube is closed by a stainless rod and thus a total reflection
boundary condition is imposed at the outlet.  Pulse waves are bounced
backward and forward in the tube multiple times before the equilibrium
state is restored.  We measured at two points, $A$ and $B$, which are
respectively close to the proximal and distal ends of the tube.
Table~\ref{tab:Param_tube_fluid} summarizes the parameters of the
elastic tube and fluid: the thickness of the wall $h$, the reference
diameter $D$, the total length of the tube $L$, the distances from the
inlet to the two measurement points $L_{A}$ and $L_{B}$, the fluid
density $\rho$ and the kinematic viscosity $\nu$.

\begin{table}[ht]
\centering
\begin{tabular}{ c | c | c | c | c | c | c}
\hline
$h~(\text{cm}) $ & $D~(\text{cm})$  & $L~(\text{cm})$ & $L_{A}~(\text{cm})$ & $L_{B}~(\text{cm})$ & $\rho~(\text{kg}/\text{cm}^3)$ & $\nu~(\text{cm}^2/\text{s})$ \\
\hline
0.2 & 0.8 & 192 & 28.3 & 168.2 & 1.050$\times10^{-3}$ & 1$\times10^{-2}$ \\
\hline
\end{tabular}
\caption{Parameters of the tube and fluid.}
\label{tab:Param_tube_fluid}
\end{table}

To evaluate independently the Young's modulus of the elastic tubes we
complete the experimental setup with a tensile device.  We prepared
two specimens of the polymer of the elastic wall to use in the tensile test
(Shimadzu EZ test).  The specimens were elongated at a rate of
0.5~$\text{m}/\text{min}$ and then released at the same rate.  We
applied the least square method (linear regression) to fit the curve
against the function $F=C_0+ ES\Delta L/L$, where $C_0$ is a constant,
$E$ is the Young's modulus, $S$ is the cross-sectional area of the
specimens and $L$ is the original length.  Dividing the fitted slope
of the curve by $S$, we can estimate experimentally the Young's
modulus as 1.92$\pm$0.06  $10^5$~Pa.

\begin{figure}
\centering
\includegraphics[width=0.75\textwidth]{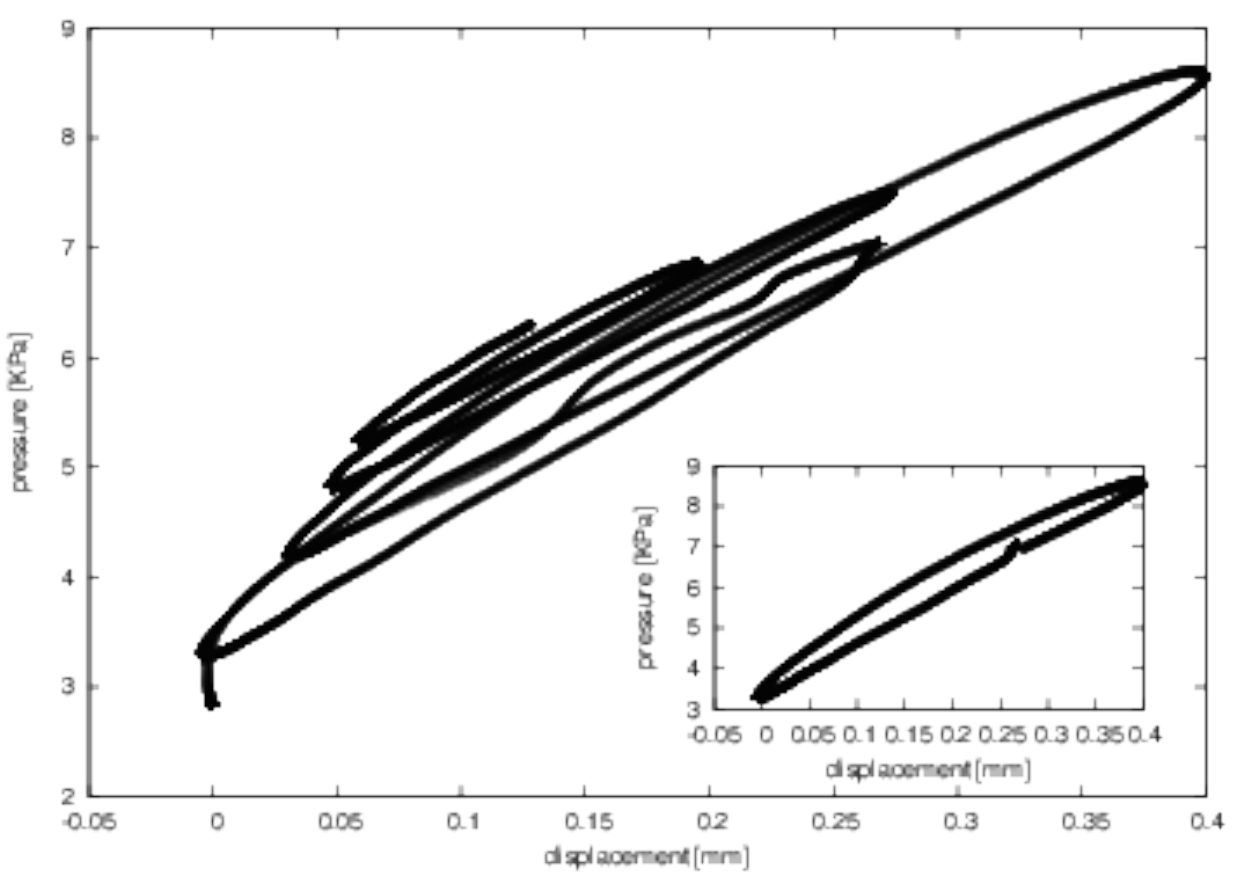}
\caption{Experimental pressure-radius (P-R) loop. Insert : one period loop.  Note that the system
  in the linear regime.}
\label{fig:p-r}
\end{figure}

\subsection{Parameter estimation}

We present the method used for the evaluation of the Young's modulus,
the wall viscosity and the fluid friction.

\subsubsection{Young's modulus}

In order to estimate the Young's modulus $E$ we propose two different
methods: using numerical simulations and by
integration of the experimental pressure-radius curve shown in
Figure~\ref{fig:p-r}. We note that the system is in the linear zone.
The values of $E$ computed in each approach will be compared to those
given by the tensile test.


\paragraph{Numerical simulations}

In the first approach, using the fact the velocity of pulse wave is
directly related to the stiffness through the Moens-Korteweg
formula~\cite{formaggia2003one}, we vary Young's modulus in
numerical simulations to match the wave peaks coming from experimental
signal taken in points $A$ and $B$. The best fit will give the optimal
Young's modulus $E_0$.

\paragraph{Integration of the experimental pressure-radius signal}

In the second approach we use the experimental data and impose a
sinusoidal wave of only one full period strictly. The net volume
of fluid injected into the tube was zero, and the tube returned to the
original state with the amplitude dampened roughly in a oscillatory
way. In this situation the energy loss is due to the wall
viscosity. Integrating the viscoelastic tube
law~(\ref{constitutive_exp}) times the wall velocity $\frac{dR}{dt}$
from the starting time $t_0$ to the final time $t_e$ we found that the
work done by the mechanical system is
\begin{equation}
\int _{t_0} ^{t_e} (P-P_{ext} ) \frac{dR}{dt} \ dt=\int _{t_0} ^{t_e}
  \nu_e(R-R_0) \frac{dR}{dt} \ dt + \int _{t_0} ^{t_e} \nu_s
  \left(\frac{dR}{dt} \right)^2 \ dt. \label{work}
\end{equation}
From the time series of the pressure $P(t)$ and the wall displacement
$R(t)$ the evaluation of the viscoelastic term $\nu_s$ is
straightforward as long as both the external pressure $P_{ext}$ and
the work done by the elastic component  (the 1st term of the rhs of
equation~(\ref{work})) are zero. Once the viscosity coefficient
$\nu_s$ is calculated, the tube law~(\ref{constitutive_exp}) can be
rearranged to give $P-P_{ext}-\nu_s (dR/dt)=\nu_e(R-R_0)$ and the
elastic coefficient $\nu_e$ can be estimated by linear regression. We
note that we have additionally estimated the viscoelastic term.

\subsubsection{Viscoelastic parameters}
\label{sec:visc-param}

For the estimation of the viscoelasticity parameters, we introduce a cost
function defined by the normalized root mean square (NRMS) error
between the experimental signal of pressure $P_{exp}$ and the numerical
predictions $P_{sim}$
\[
NRMS=\frac{1}{\max({P_{exp}})-\min({P_{exp}}) }\sqrt{\frac{\sum_N({P_{sim}-P_{exp}})^2}{N}},
\]
where $N$ is the number of temporal data points and $P_{sim}$ depends
on the fluid friction and wall viscosity for fixed Young's modulus
$E_0$. For each run we obtain numerically the temporal series of the
cross-sectional area $A$ from equation \eqref{eq:1DA} and compute the
numerical prediction of the pressure using
equation~\eqref{constitutive_exp}.  In practice, we fixed $C_f$ for
different values from $8\pi \nu$ to $33\pi \nu$, and for each value,
we fitted the parameters $\phi$ by minimizing the NRMS. As $C_f$ was
fixed for each step we only did an one dimensional minimization by
doing small variations of $\phi$ to find the minimum.  This is
particular case of the Steepest Descent approach for a functional minimum,
where the new search direction is orthogonal to the previous.  The
parameter optimisation was done on the two measurement points $A$ and
$B$, and the consistency of the results estimated from the two sets of
data was checked.

\section{Results}

In this Section we present the results of the parameter estimations
using the methods described before.  Please note
that the final state on the experimental data as well as the numerical results
has a higher pressure than the initial state.  That is because we
imposed a half sinusoidal wave at the inlet and thus a net volume of
about 4.5~$\text{cm}^3$ fluid was injected into the tube. Only in the
case when we do the integration of the experimental pressure-radius
signal to computed the wall viscosity and the fluid friction we impose
a complete period at the inlet in order of to have no net extra volume inside
the elastic tube.

\subsection{Young's modulus}

We vary Young's modulus $E$ in different simulations imposing a half
sinusoidal wave at the inlet. Numerical simulations were done for $E$
starting from $2.00\times 10^5$~Pa to $2.15\times 10^5$~Pa, with a
step of $0.01\times 10^5$~Pa.  We have found that for the value of
$E \sim E_0=2.08\times 10^5$~Pa, the difference of the arrival times
between the experimental signal and predictions at the measurements
points $A$ and $B$ was minimal (smaller than 0.02 \ s for each of the
first ten peaks). The Figure \ref{fig:fit_E} shows the variations of
the arrival times when we change the Young's modulus.

\begin{table}[ht]
\centering
\begin{tabular}{ c  c | c | c  }
\hline
  \multicolumn{2}{c|}{method}   & $E~(10^5~\text{Pa})$ & $\phi~(\text{kPa} \cdot \text{s} )$  \\
\hline
\multicolumn{2}{c|}{Numerical} & 2.08 & 1.0  \\
\hline
 \multicolumn{2}{c|}{Integration P-R data} & 1.45---2.90 &  0.97---1.94 \\
\hline
\multicolumn{2}{c|}{Tensile test} & 1.92$\pm$0.06 & -  \\
\hline
\end{tabular}
\caption{Young's modulus and Viscoelasticity of the polymer computed
  using three different approaches : 1D model optimisation,
  Pressure-radius experimental data and tensile test. }
\label{tab:Viscoelasticity}
\end{table}

This value is in the range estimated with the integrated method
$[1.45-2.9 \ 10^5]$ and is about $8\%$ bigger than those give by the
tensile device (1.92$\pm$0.06 $10^5$).  Besides the measurement error, the
variance in the home-made polymer tubes may also contribute to the
difference.

\subsection{Fluid friction and wall viscosity}

\begin{figure}
\centering
\includegraphics[width=0.75\textwidth]{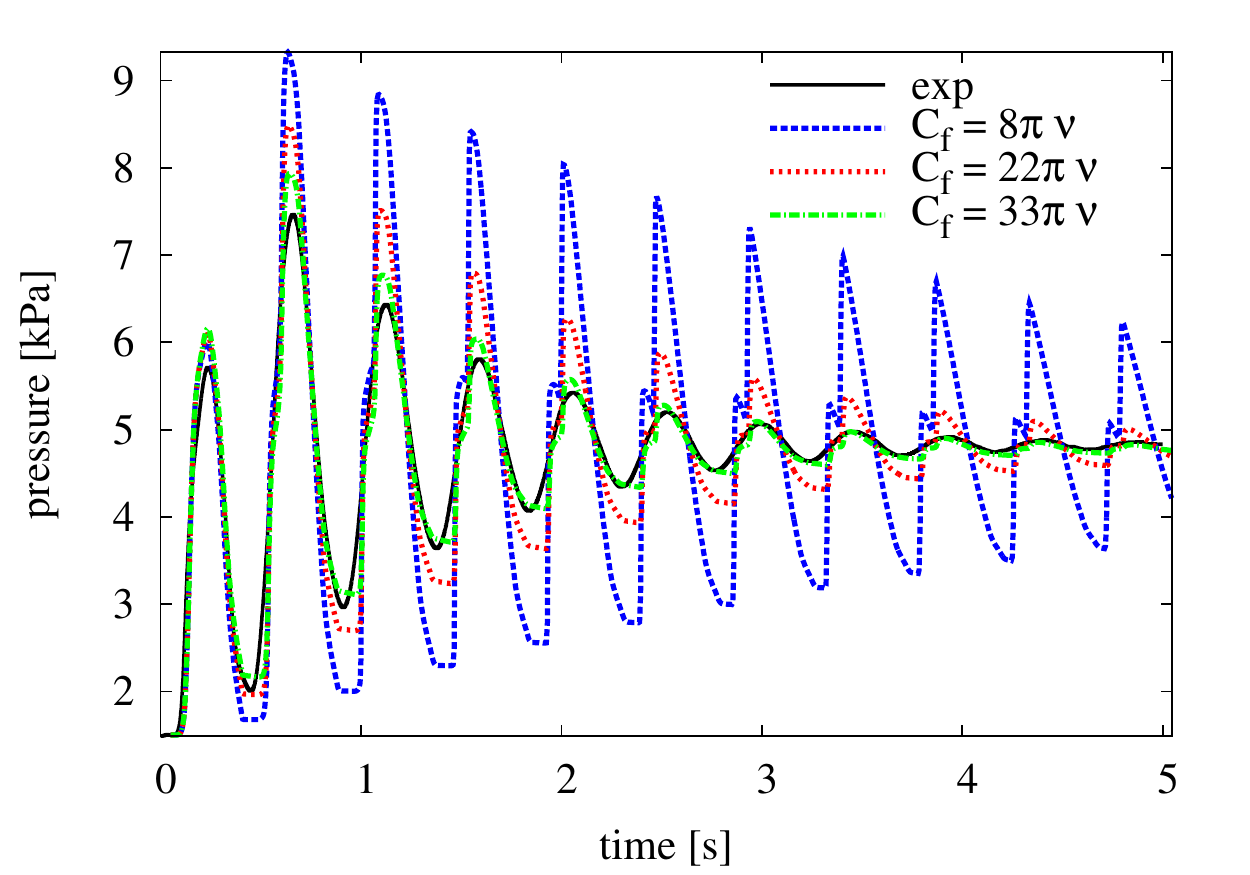}
\caption{Pressure time series at measurement point A.
The elastic model predicts shocks.
Increasing the friction term can damp the amplitude effectively, but the shocks still exist.
$E=2.08\times 10^5$~$\text{Pa}$ and $\phi=0$.}
\label{fig:fit_Cf}
\end{figure}

The friction and wall viscosity terms are both damping
factors in the model equation. The key point is to be able of
discriminate them when we are looking for the optimal values.

First we used an pure elastic model (the wall viscosity $\phi$ is set
to 0) and we
varied the friction coefficient $C_f$. Fig.~\ref{fig:fit_Cf} presents
the runs (called waves) with three values of the friction coefficient $C_f$:
$8\pi \nu$, $22\pi \nu$ and $33\pi \nu$.  Using the first value,
derived from a parabolic velocity profile, the predicted pressure wave
has two main unrealistic features: (i) we have an overestimated
pressure amplitude and (ii) we develop discontinuities or shocks, in
contradiction to the experimental measurement (blue line,
Fig.~\ref{fig:fit_Cf}).  The second value comes from Smith et
al.~\cite{smith2002anatomically}, and we can see the amplitude becomes
closer to the experimental one (red line, Fig.~\ref{fig:fit_Cf}).  The
third value gives the best prediction in terms of pressure amplitude
but there are still discontinuities or shocks (green line,
Fig.~\ref{fig:fit_Cf}). We recall that, for a pure elastic model, we
have always a finite time discontinuities, which is proper to the
hyperbolic structure of the governing equations.

\begin{table}[!ht]
\centering
\begin{tabular}{  c | c  c  c  c  c  c  c  }
\hline
 & wave1  & wave2 & wave3  & wave4 & wave5 & wave6 & wave7  \\
\hline
$C_f (\pi \nu)$ & 8 & 14 & 18 & 22 & 26 & 30 & 33   \\
\hline
$\phi (\text{kPa} \cdot \text{s} ) $ & 2.0 & 1.6 & 1.3 & 1.0 & 0.8 & 0.5 & 0.4 \\
\hline
NRMS (\%) & 1.96 & 1.75 & 1.66 & 1.64 & 1.74 & 1.92 & 2.15   \\
\hline
\end{tabular}
\caption{Parameters of fluid friction and wall viscosity and the
  corresponding NRMS. Each wave correspond to a different run.}
\label{tab:Cf_Phi}
\end{table}

Table~\ref{tab:Cf_Phi} summarizes the runs (wave 1 to 7) for different
values of $C_f$, together with the optimal value of $\phi$ found by
optimization and the
corresponding residuals of NRMS.  We observe for increasing values of
$C_f$ increases that the parameter $\phi$ decreases.  The minimal
residual of NRMS achieves for wave4 and the limit cases (wave 1 and
wave 7) are the worsts.

\begin{figure}
\centering
\subfigure[]{\label{fig:fit_Cf_Phi}\includegraphics[width=0.7\textwidth]{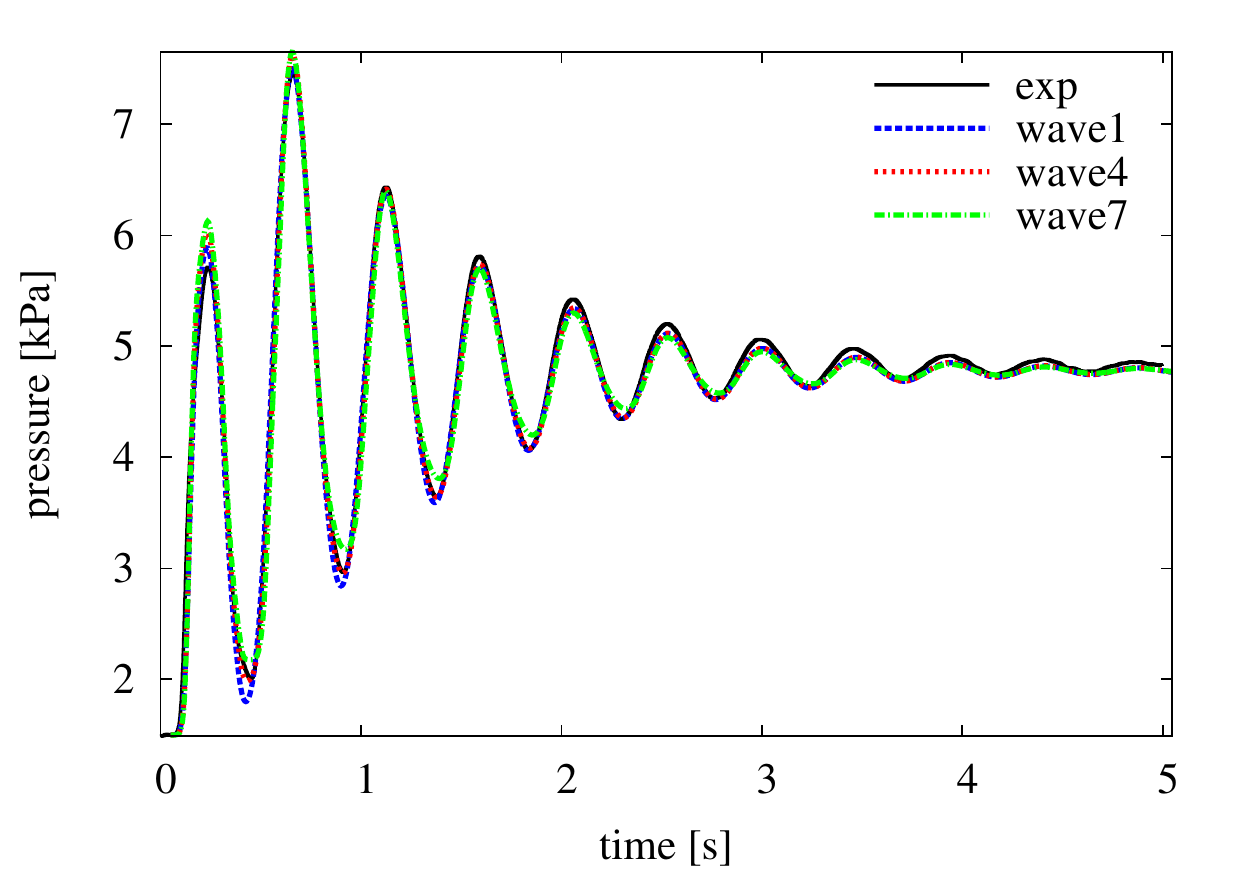}}

\subfigure[]{\label{fig:spectrum_pressure}\includegraphics[width=0.7\textwidth]{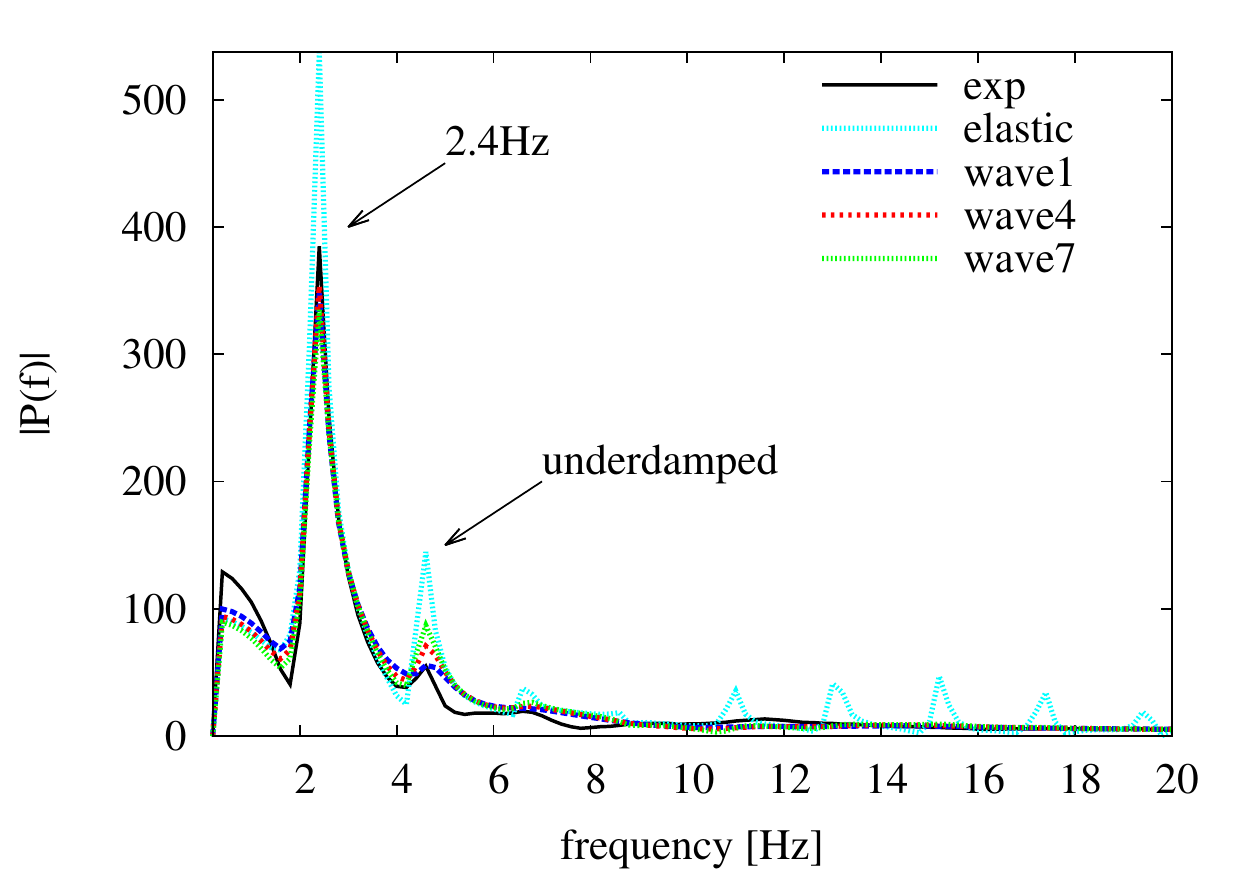} }
\caption{Experiments (line labelled exp) and simulations at measurement point $A$. Left
  (a): pressure time series.
Right (b): spectrum of the pressure series (only frequencies less than 20 Hz are shown).
$E=2.08\times 10^5$~Pa.
For the elastic case, $C_f=22\pi \nu$ and $\phi=0$.
The values of $C_f$ and $\phi$ for the three viscoelastic waves are shown in Table~\ref{tab:Cf_Phi}. }
\label{fig:fit_Cf_Phi_two}
\end{figure}

We plotted waves 1, 4 and 7 in Fig.~\ref{fig:fit_Cf_Phi}.  First we
noticed that the discontinuities or shocks disappear and that the
amplitude of the three waves are close to the experimental data.
However, in the first two seconds of the temporal series, the
wave-front of wave 7 is
steeper than the others.  This difference is more clear when we plot the
power spectrum of the time series (Fig.~\ref{fig:spectrum_pressure}),
which shows that the high frequency components of wave 7 are
underdamped.  This is because the damping effect of wall viscosity is
stronger on high frequency waves while that the fluid friction does
not depend on the frequency in our model.  In the last part of the
time records, only the main harmonic is still present, thus the
difference between the three simulated waves is very small.  The
viscoelastic parameters estimated by the presented methods are
summarized in Table~\ref{tab:Viscoelasticity}.  The values estimated
by the data fitting with the 1D model fall into the range measured by
the integrated approach of the pressure-radius (P-R) series data.

\subsection{Sensitivity study}
\label{sec:sensibility-study}

\begin{figure}
\centering
\includegraphics[width=0.75\textwidth]{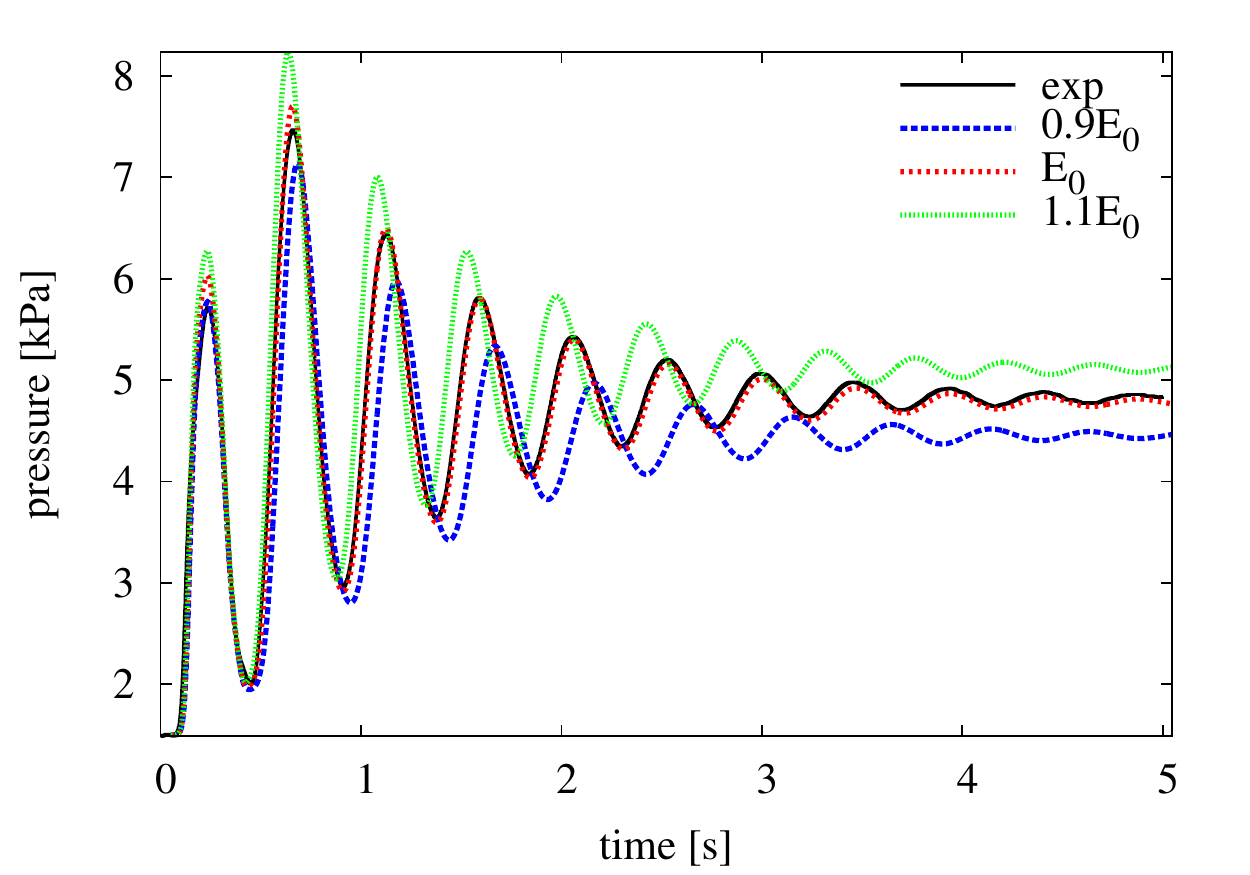}
\caption{Sensitivity study. Pressure time series at measurement point
  A. $E_0$ is the best fit for the Young's modulus.
 If $E_0$ is perturbed 10\%, the arrival time of each peak changes significantly.
$C_f=22\pi \nu$ and $\phi=0.9$~$\text{kPa} \cdot \text{s}$.}
\label{fig:fit_E}
\end{figure}

Fig.~\ref{fig:fit_E} presents the parameter sensitivity
for Young's modulus $E$  having a variance of 10\%
around $E_0$.  The arrival time of each peak is significantly later
when $E$ decreases and vice versa.

\begin{figure}
\centering
\begin{tabular}{c c}

\subfigure[]{\label{fig:Cf_sensitive}\includegraphics[width=0.48\textwidth]{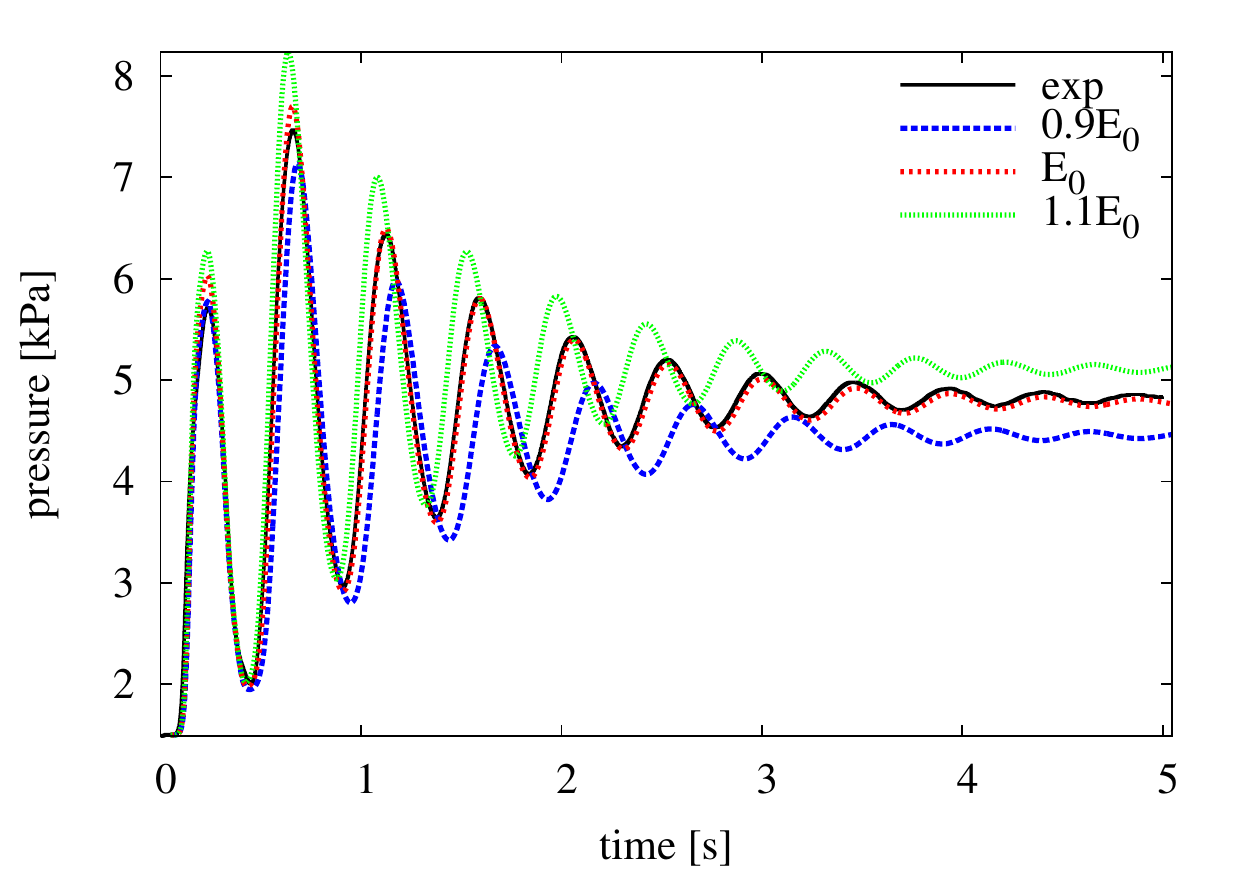} }
&\subfigure[]{\label{fig:Phi_sensitive}\includegraphics[width=0.48\textwidth]{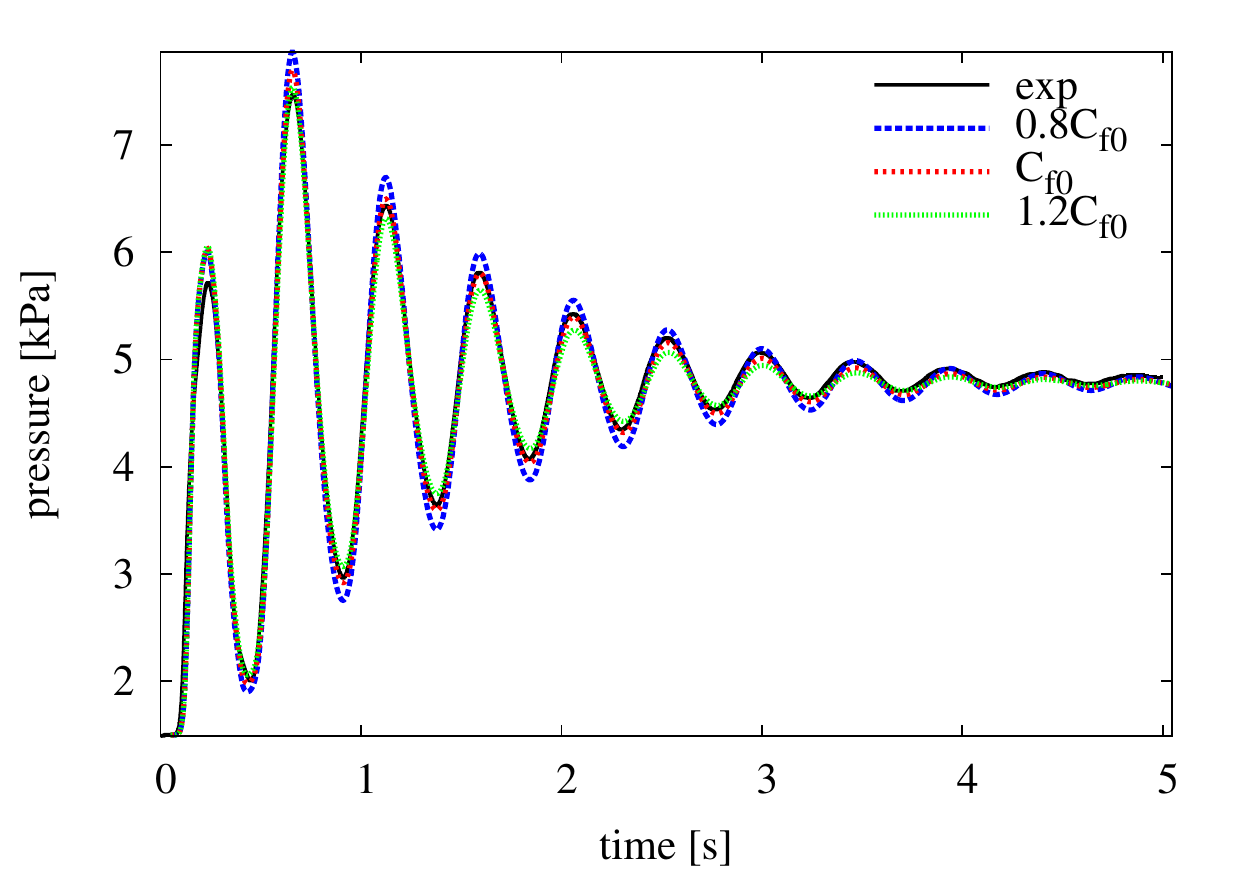}} \\
\end{tabular}
\caption{Time series of pressure with a 20\% uncertainty of $C_f$ (left) and $\phi$ (right).}
\label{fig:sensitive}
\end{figure}

We also tested the sensitivity of the model to $C_f$, $\phi$ and $\alpha$.
For $C_f$ and $\phi$, an uncertainty of about 20\% produces a moderate
variance on the predicted wave (see Fig.~\ref{fig:Cf_sensitive} and
~\ref{fig:Phi_sensitive}).  The sensitivity of the output to $C_f$ and
$\phi$ is in the same order.  In contrast, when $\alpha$ is tested in
the range from 1.0 to 1.3, there is no noticeable difference between
the numerical predictions.  Thus, the value of $\alpha$ can be set to
1.0.  There exists indeed more sophisticated sensitivity techniques
\cite{xiu2007parametric}  but it is  beyond the presented study.

\subsection{Integration schemes}
\label{sec:integration-schemes}

\begin{figure}
\centering
\begin{tabular}{ c c}
Measurement $A$ & 	Measurement $B$ \\

\subfigure[]{\label{fig:without viscosity}\includegraphics[width=0.48\textwidth]{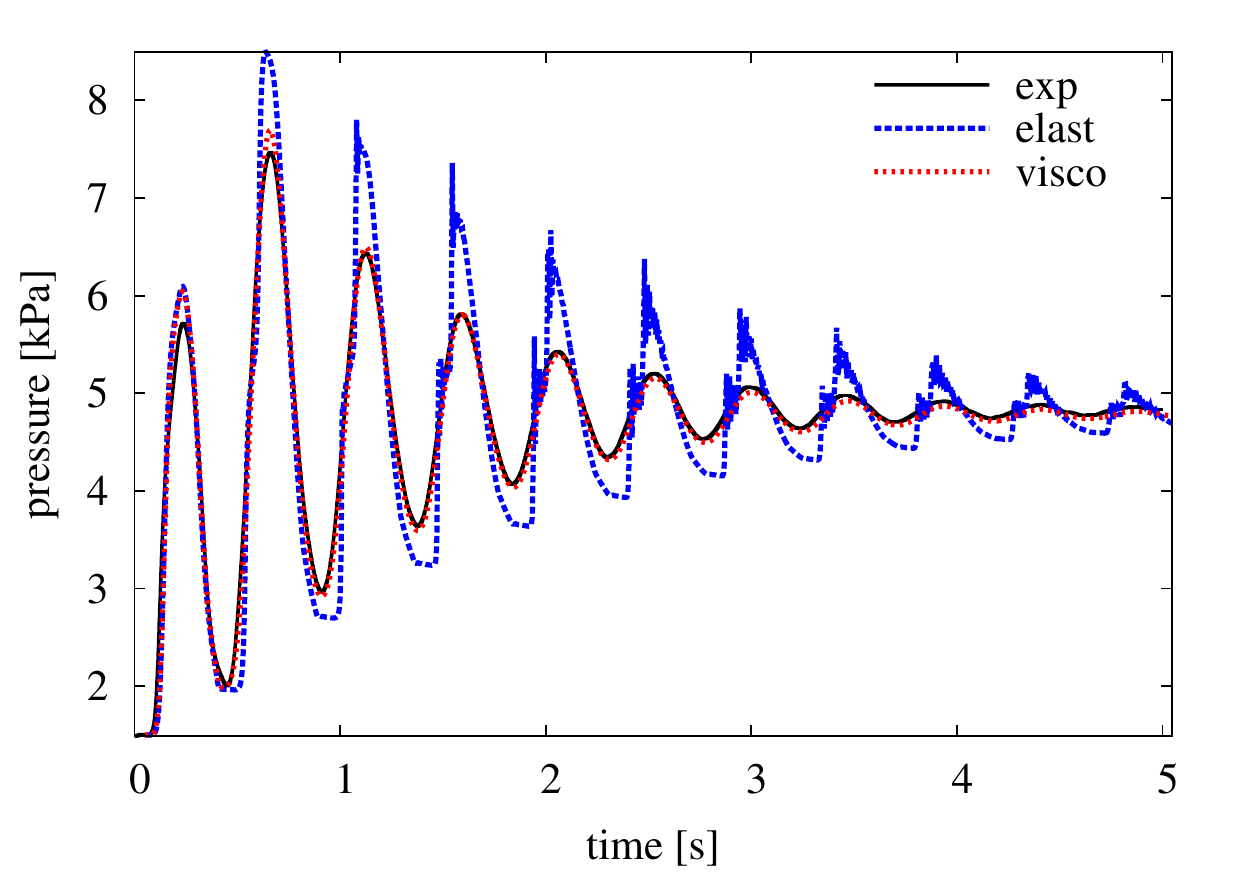} }
&\subfigure[]{\label{fig:with viscosity}\includegraphics[width=0.48\textwidth]{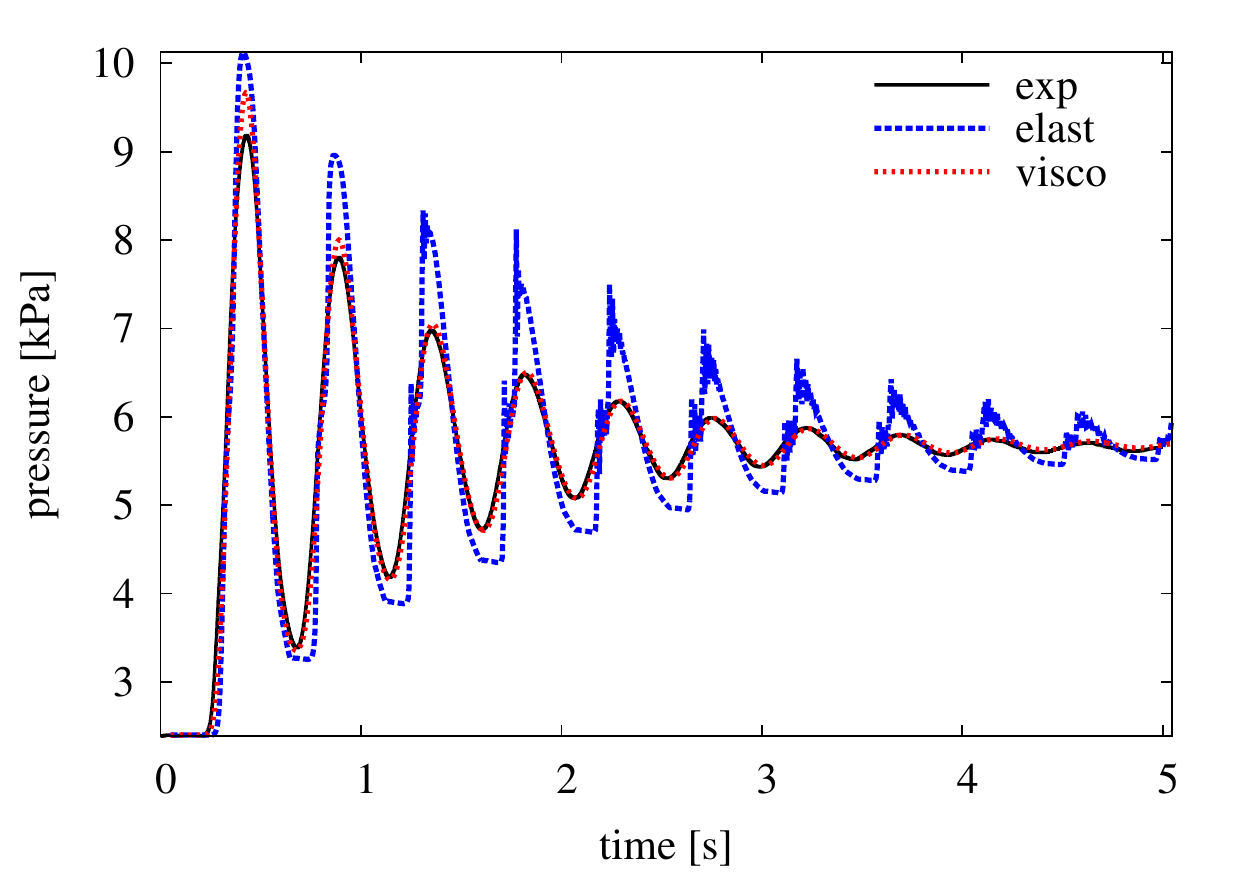}} \\
\subfigure[]{\label{fig:without viscosity}\includegraphics[width=0.48\textwidth]{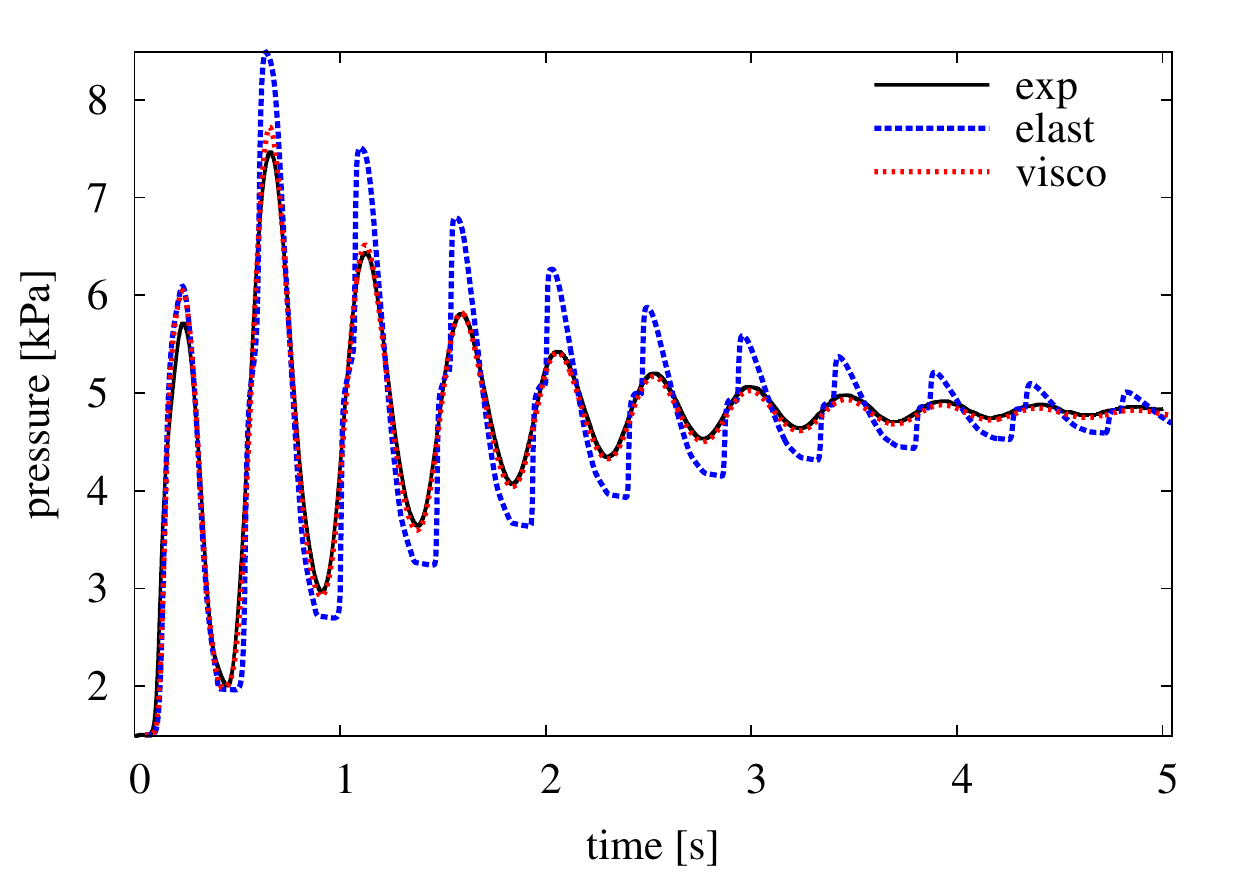} }
&\subfigure[]{\label{fig:with viscosity}\includegraphics[width=0.48\textwidth]{fig7c.pdf}} \\
\end{tabular}
\caption{Pressure time series at the two measurement points with two numerical schemes.
Left column: point $A$, right column: point $B$. Top row: MacCormack method, bottom row: MUSCL method.
The viscoelastic model predicts much better than the elastic model at both the measurement points.
The MUSCL method depresses the numerical oscillations when there are shocks.
The parameters are: $E=2.08\times 10^5$~Pa, $C_f=22\pi \nu$, and $\phi$=1.0~$\text{kPa}\cdot\text{s}$ (visco).}
\label{fig:compare_total}
\end{figure}

We tested two different integration schemes : MacCormack and MUSCL. We
compared the performances for a pure elastic as well as for a
viscoelastic model.  In Fig.~\ref{fig:compare_total}, we plotted the
pressure waves for the numerical predictions against the experiments
data at the two measurement points: left column for point $A$ and
right column for point $B$.

The discontinuities or shocks predicted by the elastic model are very
obvious.  The MacCormack scheme produces numerical oscillations (top
row) whereas the MUSCL scheme depresses them because it includes a
slope limiter (bottom row).  For the viscoelastic model, the shocks
disappear and a much better agreement is found at both
locations $A$ and $B$.  If the solution is quite smooth, there is essentially no
difference between the two numerical schemes in accuracy.  The
consistency between the two locations make us
confident in the agreement between experiments and numerical simulations.

\section{Discussion}

We evaluated the  stiffness and friction within  a nonlinear
1D fluid dynamical model with a viscoelastic law for the wall
mechanics against experimental data.

The value of vessel stiffness estimated by the 1D model was compared
to values measured using a
tensile test.  We notethat a small variance in stiffness can
significantly the change the mean pressure, pulse pressure and wave
velocity. Under the operating pressure within our experiment, the
nonlinearity seems not large as shown in Figure~\ref{fig:p-r}.
However, we note that the nonlinearity may be more significant under
physiological conditions.  More studies have to be done to evaluate the
nonlinear elasticity of the arteries under real conditions.

The fluid friction and wall viscosity were fitted from experimental data using the 1D model.
We obtained good agreement between the 1D model results and
experiments.  If experimental uncertainties are considered, it can be
estimated that $C_f=22\pm 4\pi \nu$ and
$\phi=1.0\pm0.3~\text{kPa}\cdot \text{s}$ (determined by the runs  wave3 and
wave5 in Table~\ref{tab:Viscoelasticity}).
Our results confirm that in cases of blood flow with a similar
characteristic Womersley number, the Poiseuille model underestimates
the fluid friction (see e.g.~\cite{saito2011one}).  The widely used
value $C_f=22\pi \nu$ in large arteries is then acceptable.  However, in
smaller arteries, the Womersley number can be less than one, so a
parabolic velocity profile is more likely to appear, which implies
that $C_f$ decreases to $8\pi \nu$.  Thus the friction term should
vary through the whole cardiovascular system and a smaller value of
$C_f$ should be considered if the Womersley number is smaller.

 In our experimental study,
the frequency of the main harmonic is 2.4 Hz (see
Fig.~\ref{fig:spectrum_pressure}) and thus the Womersley number is
about 15.5.  This value is only slightly bigger than the Womersley
number at the ascending aorta which is
13.2~\cite{fung1997biomechanics}.
  Under \textit{in vivo} conditions, the wall viscosity is much
larger as measured by Armentano et al.~\cite{armentano1995arterial}.
However the surrounding tissues of the vessel such as fat may also damp
the waves attributed wall viscosity.  The viscoelasticity of the arteries is
mostly attributed to the collagen and elastin fibers in the wall,
which is different from the polymer tube.

The viscoelasticity of the wall dampens the high frequency components
of the wave, thus the waveform is not very front-steepened, which has
been pointed out by many previous studies (see
e.g.~\cite{alastruey2011pulse,holenstein1980viscoelastic}).  A
perturbation of 20\% on wall viscosity introduces moderate variances
on the pressure waveform, which is similar to the fluid friction (see
Fig.~\ref{fig:Cf_sensitive} and ~\ref{fig:Phi_sensitive}).  The output
of the 1D model is not very sensitive to uncertainties of the two
damping factors.  Thus it is possible to use general values of those
two parameters even in patient-specific simulations with the 1D model.

We solved the nonlinear 1D viscoelastic model with MacCormack and
MUSCL schemes.  The elastic model predicts shocks, which are captured
by the MUSCL method without non-physical oscillations.

Some limitations of our approach are : while the flow
rate may be similar, material properties are likely different and the
in vivo (invasive) pressure measurements could hardly to including  in a clinical
protocol. One could advance that in real arteries under normal
physiological conditions, discontinuities or shocks are not present but
in pathologocal situations (anasthomoses, artheromes) or after
surgeries (i.e. stent) the discontinuities on the Young's modulus of the
arterial wall  can lead to flow discontinuities. Concerning the
boundary conditions, arteries never display this type of vessel
ending but it is not unreasonable to image a clinical protocol with
a short stopping blood flow to observe localized backward waves.

\section{Conclusion}

We studied and evaluated the parameters of the nonlinear 1D
viscoelastic model using data from an experimental setup.  The 1D
model was solved by two schemes, one of which is shock-capturing.

The value of vessel stiffness, estimated by the 1D model was
consistent with values obtained by an integrated method using
experimental data (pressure-radius time series) and tensile tests.
The fluid
friction and wall viscosity were fitted from data measured at
two different locations.  The estimated viscoelasticity parameters were
consistent with values obtained with other methods.  The good
agreement between the predictions and the experiments indicate that
the nonlinear 1D viscoelastic model can simulate the pulsatile blood
flow very well.  We showed that the effect of wall viscosity on the
pulse wave is as important as that of fluid viscosity.

\section*{Acknowledgements}
This work was supported by French state funds managed by CALSIMLAB and
the ANR within the investissements d'Avenir programme under reference
ANR-11-IDEX-0004-02.  We are very grateful to the anonymous reviewers,
whose comments helped us a lot to improve this paper.

\section*{Conflict of interest}
All the authors have been involved in the design of the study and the
interpretation of the data and they concur with its content. There are
no conflicts of interest between the authors of this paper and other
external researchers or organizations that could have inappropriately
influenced this work.
\bibliographystyle{plain}
\bibliography{Numerical}
\end{document}